\documentclass[11pt,epsf]{article}
\usepackage{hyperref}
\usepackage{subfigure}
\usepackage{graphics}
\usepackage{amssymb}
\usepackage{amsmath}
\usepackage{array}
\usepackage{rotating}
\usepackage{axodraw}
\usepackage{pstricks}

\def\d{\delta}

\def\eps{\epsilon}
\def\a{\alpha}
\def\b{\beta}

\def\g{\gamma}

\def\si{\sigma}
\def\lag{\langle}
\def\rag{\rangle}

\newcommand{\sq}{\square}
\newcommand{\bea}{\begin{eqnarray}}
\newcommand{\eea}{\end{eqnarray}}
\newcommand{\bann}{\begin{eqnarray*}}
\newcommand{\eann}{\end{eqnarray*}}
\newcommand{\bmi}{\begin{minipage}}
\newcommand{\emi}{\end{minipage}}

\newcommand{\psl}{p \! \! \!  /}
\newcommand{\qsl}{q \! \! \!  /}
\newcommand{\lsl}{l \! \! \!  /}

\newcommand{\beqa}{\begin{eqnarray}}
\newcommand{\eeqa}{\end{eqnarray}}
\def\beq{\begin{equation}}
\def\eeq{\end{equation}}
\def\nn{\nonumber}

\begin{document}

\begin{center}
\vspace{.5cm}

{\bf\large  Trace Anomaly, Massless Scalars\\ and the Gravitational Coupling of QCD}

\vspace{1.5cm}
{\bf\large Roberta Armillis, Claudio Corian\`{o}, Luigi Delle Rose
\footnote{roberta.armillis@le.infn.it, claudio.coriano@le.infn.it, luigi.dellerose@le.infn.it}}
\vspace{1cm}

{\it  Dipartimento di Fisica, Universit\`{a} del Salento \\
and  INFN Sezione di Lecce,  Via Arnesano 73100 Lecce, Italy}\\
\vspace{.5cm}

\begin{abstract}
The anomalous effective action describing the coupling of gravity to a non-abelian gauge theory can be determined by a variational solution of the anomaly equation, as shown by Riegert long ago. It is given by a nonlocal expression, with the
nonlocal interaction determined by the Green's function of a conformally covariant operator of fourth order. In recent works it has been shown that this interaction is mediated by a simple pole in an expansion around a Minkowski background,  coupled in the infrared in the massless fermion limit.  This result relies on the local formulation of the original action in terms of two auxiliary fields, one physical scalar and one ghost, which take the role of massless composite degrees of freedom. 
In the gravity case, the two scalars have provided ground in favour of some recent proposals of an infrared approach to the solution of the dark energy problem, entirely based on the behaviour of the vacuum energy at the QCD phase transition. 
As a test of this general result, we perform a complete one-loop computation of the effective action describing the coupling of a non-abelian gauge theory to gravity. We confirm the appearance of an anomaly pole which contributes to the trace part of the $TJJ$ correlator and of extra poles in its trace-free part, in the quark and gluon sectors, describing the coupling of the energy momentum tensor ($T$) to two non abelian gauge currents ($J$).  

\end{abstract}
\end{center}
\newpage

\section{Introduction}
The study of the effective action describing the coupling of a gauge theory to gravity via the trace anomaly \cite{Duff:1977ay} is an important aspect of quantum field theory, which is not deprived also of direct phenomenological implications. This coupling is mediated by the correlator involving the energy momentum tensor together with two vector currents (or $TJJ$ vertex), which describes the interaction of a graviton with two photons or two gluons in QED and QCD, respectively.
At the same time, the vertex has been at the center of an interesting case study
of the renormalization properties of  composite operators in Yang Mills theories \cite{Collins:1994ee}, in the context of an explicit check of the violation of the Joglekar-Lee theorem \cite{Joglekar:1975nu} on the vanishing of S-matrix elements of BRST exact  operators.  In this second case it was computed on-shell, but only at zero momentum transfer.
In this work we are going to extend this computation and investigate the presence of massless singularities in its expression. These contribute to the trace anomaly and play a leading role in fixing the structure of the effective action that couples QCD to gravity. The analysis of \cite{Collins:1994ee}, which predates our study, unfortunately does not resolve the issue about the presence or the absence of the anomaly pole in the anomalous effective action of QCD
because of the restricted kinematics involved in that analysis of the $TJJ$ vertex, and for this reason we have to proceed with a complete re-computation.

Anomaly poles characterize quite universally (gravitational and chiral) anomalous effective actions, in the sense that account for their anomalies. They have been identified and discussed in the abelian case both by a dispersive analysis \cite{Giannotti:2008cv} and by a direct explicit computation of the related anomalous Feynman amplitudes quite recently \cite{Armillis:2009pq, Armillis:2009im}. Extensive analysis in the case of chiral gauge theory for anomalous $U(1)$ models have shown the close parallel between solutions of the Ward identities, the coupling of the poles in the ultraviolet and in the infrared region and the gravity case \cite{Armillis:2008bg, Armillis:2009sm}.

It is therefore important to check whether similar contributions appear also in non-abelian gauge theories coupled to gravity. We recall that the same pole structure is found in the variational solution of the expression of the trace anomaly, where one tries to identify an action whose energy momentum tensor reproduces the trace anomaly. This action, found by Riegert long ago \cite{Riegert:1984kt}, is nonlocal and involves the Green function of a quartic (conformally covariant) operator. The action describes the structure of the singularities of anomalous correlators with any number of insertions of the energy momentum tensor and two photons ($T^n JJ$), which are expected to correspond both to single
and to higher order poles, for a sufficiently high $n$. For obvious reasons, explicit checks of this effective action using perturbation theory - as the number of external graviton lines grows -  becomes increasingly difficult to handle. The $TJJ$ correlator is the first (leading) contribution  to this infinite  sum of correlators in which the anomalous gravitational effective action is expanded.

Given the presence of a quartic operator in Riegert's nonlocal action, the proof that this action contains a single pole to lowest order (in the TJJ vertex), once expanded around flat space, has been given in \cite{Giannotti:2008cv} by Giannotti and Mottola, and provides the basis for the discussion of the anomalous effective action in terms of massless auxiliary fields contained in their work. The auxiliary fields are introduced in order to rewrite the action in a local form. We show by an explicit computation of the lowest order vertex that Riegert's action is indeed consistent in the non-abelian case as well, since its pole structure is recovered in perturbation theory, similarly to the abelian case. Therefore, one can reasonably conjecture the presence of anomaly poles in each gauge invariant subsets of the diagrammatic expansion, as the computation for the non-abelian TJJ shows (here for the case of the single pole). In particular, this is in agreement with the result of \cite{Giannotti:2008cv},
where it is shown that, after expanding around flat spacetime, the quartic operator in Riegert's action becomes a simple $1/\square$ nonlocal interaction (for the $TJJ$ contribution), i.e. a pole term. We remark that the identification of a pole term in this and in others similar correlators, as we are going to emphasize in the following sections 
(at least in the case of QED and for the sector of QCD mediated by quark loops), requires an extrapolation to the massless fermion limit, and for this reason its interpretation as a long-range dynamical effect in the gravitational effective action requires some caution. In QCD, however, there is an extra sector that contributes to the 
same correlator, entirely due to virtual loops of gluons in the anomaly graphs, which remains unaffected by the massless fermion limit. The appearance of such a singularity in the effective action, however, does not necessarily imply that its contribution survives in the physical S-matrix. We will also establish the appearance of other singularities in the trace-free 
form factors which, obviously, are not part of Riegert's action. 

We will comment in our conclusions on the possible implications of these results and on some recent proposals to link this type of behaviour \cite{Urban:2009vy, Urban:2009yg} to cosmology and to the dark energy problem. We also remark that, in general, the coefficient in front of the trace anomaly, for a given theory, can be computed in terms of its massless fields content, and as such it is well known. However, the structure of the effective action and the characterization of its fundamental form factors at nonzero momentum transfer and its complete analytical structure is a novel result. In this respect, the classification of all the
relevant tensor structures which appear in the computation of this correlator is rather involved and has been performed
in the completely off-shell case.  We remark that the complexity of the final expression, in the off-shell case, prevents us from presenting
its form. For this reason we will give only the on-shell version of the complete vertex, which is expressed, as we have mentioned, only in terms of three fundamental form factors.

Concerning the phenomenological relevance of this vertex, we just mention that it plays an essential role in the study of NLO corrections to processes involving a graviton exchange. In fact, in theories with extra dimension, where a low-gravity scale and the presence of Kaluza-Klein excitations may enhance the rates for processes mediated by gluons and gravitons, the vertex appears in the hard scattering of the corresponding factorization formula \cite{Mathews:2004xp} and has been computed in dimensional regularization. However, to our knowledge,  in all cases, there has been no separate discussion of the general structure of the vertex (i.e. as an amplitude) nor of its Ward identities, which, in principle, would require a more careful investigation because of the trace anomaly. Anomalous amplitudes, in fact, are defined by the fundamental Ward identities imposed on them, that we are going to derive from general principles. We cover this gap and show, that both dimensional regularization and dimensional reduction reproduce the correct Ward identity satisfied by this vertex, showing at the same time that the use of these regularizations is indeed appropriate. Results for this vertex will be given only in the on-shell case, since in this case the result can be expressed in terms of just three form factors. We have computed also the off-shell effective action,
but its expression is rather lengthy and will not be discussed here, since it is gauge dependent and of less significance compared to the on-shell result. Most of our work is concerned with a technical derivation of the leading contribution to the anomalous effective action of QCD coupled to gravity. We have summarized in our conclusions a brief discussion of the relevance of this study in the ongoing attempt to link the trace anomaly and QCD to a possible alternative solution of the problem of
dark energy, using this effective action as an intermediate step  \cite{Urban:2009vy, Urban:2009yg}.

\section{Anomalous effective actions and their variational solutions}
In this section we briefly review the topic of the variational solutions of anomalous effective actions, and on the local formulations of these using auxiliary fields.
\begin{figure}[t]
\begin{center}
\includegraphics[scale=1.0]{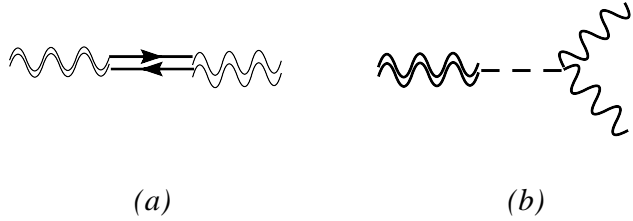}
\caption{\small The diagrams describing the anomaly pole in the dispersive approach. Fig. (a) depicts the singularity of the spectral density $\rho(s)$ as a spacetime process. Fig. (b) describes the anomalous pole part of the interaction via the exchange of a pole.}
\label{collinear}
\end{center}
\end{figure}

One well known result of quantum gravity is that the effective action of the trace anomaly is given by a nonlocal form when expressed in terms of the spacetime metric $g_{\mu\nu}$. This was obtained \cite{Riegert:1984kt} from a variational solution of the equation for the trace anomaly \cite{Duff:1977ay}
\bea
T^\mu_\mu =   b \, F + b^{\prime} \, \left( E - \frac{2}{3} \, \square \, R\right) + b'' \, \square \, R +  c\, F^{a \, \mu \nu} F^a_{\mu \nu},
\label{var2}
\eea
(see also \cite{Deser:1999zv, Deser:1993yx}
for an analysis of the gravitational sector)
which takes in $D=4$ spacetime dimensions the form
\bea
&& \hspace{-.6cm}S_{anom}[g,A] = \label{Tnonl}\\
&&\frac {1}{8}\int d^4x\sqrt{-g}\int d^4x'\sqrt{-g'} \left(E - \frac{2}{3} \square R\right)_x
 \Delta_4^{-1} (x,x')\left[ 2b\,F
 + b' \left(E - \frac{2}{3} \square R\right) + 2\, c\, F_{\mu\nu}F^{\mu\nu}\right]_{x'}. \nonumber
 \label{var1}
\eea
Here, the parameters $b$ and $b'$ are the coefficients of the Weyl tensor squared,
\beq
F = C_{\lambda\mu\nu\rho}C^{\lambda\mu\nu\rho} = R_{\lambda\mu\nu\rho}R^{\lambda\mu\nu\rho}
-2 R_{\mu\nu}R^{\mu\nu}  + \frac{R^2}{3}
\eeq
 and the Euler density
\beq
E = ^*\hskip-.2cmR_{\lambda\mu\nu\rho}\,^*\hskip-.1cm R^{\lambda\mu\nu\rho} =
R_{\lambda\mu\nu\rho}R^{\lambda\mu\nu\rho} - 4R_{\mu\nu}R^{\mu\nu}+ R^2
\eeq
 respectively
of the trace anomaly in a general background curved spacetime. Notice that the last term in (\ref{Tnonl}) is the contribution generated in the presence of a background gauge field, with coefficient $c$.
For a Dirac fermion in a classical gravitational ($g_{\mu\nu}$) and abelian ($A_{\alpha}$) background, the values of the coefficients are $b = 1/(320\,\pi^2)$, and $b' = - 11/(5760\,\pi^2)$,
and $c= -e^2/(24\,\pi^2)$, with $e$ being the electric charge of the fermion. One crucial feature of this solution is its origin,
which is purely variational. Obtained by Riegert long ago, the action was derived by solving  the variational equation satisfied by the trace of the energy momentum tensor.
$\Delta_4^{-1}(x,x')$ denotes the Green's function inverse of the
conformally covariant differential operator of fourth order, defined by
\beq
\Delta_4 \equiv  \nabla_\mu\left(\nabla^\mu\nabla^\nu + 2 R^{\mu\nu} - \frac{2}{3} R g^{\mu\nu}\right)
\nabla_\nu = \square^2 + 2 R^{\mu\nu}\nabla_\mu\nabla_\nu +\frac{1}{3} (\nabla^\mu R)
\nabla_\mu - \frac{2}{3} R \square\,.
\label{Deldef}
\eeq
Given a solution of a variational equation, it is mandatory to check whether the solution is indeed justified by a perturbative computation. One specific feature of these solutions is the presence of anomaly poles. In previous works we have elaborated on the significance of these interactions,  extracted from a direct perturbative computation, by a painstaking analysis of anomaly graphs under general kinematical conditions, and not just by a dispersive approach. The dispersive approach allows to connect this behaviour of the spectral density to a very specific infrared configuration.
\subsection{The kinematics of an anomaly pole}
In our conventions we will denote with $p$ and $q$ the outgoing momenta of the two photons/gluons and with $k$ the incoming momentum of the graviton.  $s\equiv(p + q)^2$ denotes the invariant mass of the external graviton line. A computation of the spectral density $\rho(s)$ of the $TJJ$ amplitude in QED shows that this takes the form $\rho(s)\sim \delta(s)$. The configuration responsible for the appearance of a pole is illustrated in Fig.~\ref{collinear} (a). It describes the decay of a graviton line into two on-shell photons. The decay is mediated by a collinear and on-shell fermion-antifermion pair and can be interpreted as a spacetime process. The corresponding interaction vertex, described as the exchange of a pole, is instead shown in Fig.~\ref{collinear} (b). The actual process depicted in Fig.~\ref{collinear} (a) is obtained at diagrammatic level by setting on-shell the
fermion/antifermion pair attached to the graviton line. This configuration, present in the spectral density of the diagram only for on-shell photons, generates a pole contribution which can be shown to be coupled in the infrared. This means that if we compute the residue of the amplitude for $s\to 0$ we find that it is non-vanishing. In the general expression of the vertex, a similar configuration is extracted in the high energy limit, not by a dispersive analysis, but by an explicit (off-shell) computation of the diagrams. Clearly, the pole, in this second case, has a vanishing residue as $s\to 0$, but is nevertheless a signature of the anomaly at high energy.
Either for virtual or for real photons, a direct computation of the vertex allows to extract the pole term, without having to rely on a dispersive analysis. This point has been illustrated in our previous computations of the chiral anomaly vertex \cite{Armillis:2009sm} and in the computation of the TJJ vertex for QED \cite{Armillis:2009pq}. The identification of this singularity in the case of QCD is in perfect agreement with those previous results.

\subsection{The single pole from $\Delta_4$}
In the case of the gravitational effective action, the appearance of the inverse of $\Delta_4$ operator seems to be hard to reconcile with the simpler $1/\square$ interaction which is predicted by the perturbative analysis of the $TJJ$ correlator, which manifests a single anomaly pole. In  \cite{Giannotti:2008cv}, Giannotti and Mottola show step by step how a single pole emerges from this quartic operator, by using the auxiliary field formulation of the same effective action.  Clearly, more computations are needed in order to show that the nonlocal effective action consistently does justice of {\em all} the poles
(of second order and higher) which should be present in the perturbative expansion. Obviously, the perturbative computations - being either based on dispersion theory or on complete evaluations of the vertices, as in our case - become rather hard as we increase the number of
external lines of the corresponding perturbative correlator. For instance, this check becomes almost impossible for correlators of the form $TTTT$ or higher, due to the appearance of a very large number of tensor structure in the reduction to scalar form of the tensor Feynman integrals. In the case of $TJJ$ the computation is still manageable, since it does not require Feynman integrals beyond rank-4.

Expanding around flat space, the local formulation of Riegert's action, as shown in \cite{Giannotti:2008cv,Mottola:2006ew}, can be rewritten in the form
\beq
S_{anom}[g,A]  \rightarrow  -\frac{c}{6}\int d^4x\sqrt{-g}\int d^4x'\sqrt{-g'}\, R_x
\, \square^{-1}_{x,x'}\, [F_{\alpha\beta}F^{\alpha\beta}]_{x'}\,,
\label{SSimple}
 \eeq
which is valid to first order in the fluctuation of the metric around a flat background, denoted as $h_{\mu\nu}$
\beq
g_{\mu\nu}= \eta_{\mu\nu} +\kappa h_{\mu\nu}, \qquad\qquad \kappa=\sqrt{16 \pi G_N}
\eeq
with $G_N$ being the 4-dimensional Newton's constant.
 The formulation in terms of auxiliary fields of this axion gives

\beq
S_{anom} [g,A;\varphi,\psi'] =  \int\,d^4x\,\sqrt{-g}
\left[ -\psi'\sq\,\varphi - \frac{R}{3}\, \psi'  + \frac{c}{2} F_{\alpha\beta}F^{\alpha\beta} \varphi\right]\,,
\label{effact}
\eeq
where $\phi$ and $\psi$ are the auxiliary scalar fields. They satisfy the equations
\bea
&&\psi' \equiv  b\, \sq\, \psi\,, \label{diffeq}\\
&&\square\,\psi' =  \frac{c}{2}\, F_{\alpha\beta}F^{\alpha\beta} \,,\\
&&\square\, \varphi = -\frac{R}{3}\,.
\eea

In order to make contact with the $TJJ$ amplitude, one needs the expression of the energy momentum extracted from
(\ref{effact}) to leading order in $h_{\mu\nu}$, or, equivalently, from (\ref{SSimple}) that can be shown to take the form

\beq
T^{\mu\nu}_{anom}(z) =
\frac{c}{3}  \left(g^{\mu\nu}\sq - \partial^{\mu}\partial^{\nu}\right)_z \int\,d^4x'\, \sq_{z,x'}^{-1}
\left[F_{\alpha\beta}F^{\alpha\beta}\right]_{x'}.
\label{Tanom}
\eeq
Notice that $T^{\mu\nu}_{anom}$ is the expression of the energy momentum tensor of the theory in the background of the gravitational and gauge fields. We recall, in fact, that in the QED case, for instance, the energy momentum tensor of the theory is split into the free fermionic part $T_f$,  the interacting fermion-photon part
$T_{fp}$ and the photon contribution $T_{ph}$ which are given by
\beq
T^{\mu\nu}_{f} = -i \bar\psi \gamma^{(\mu}\!\!
\stackrel{\leftrightarrow}{\partial}\!^{\nu)}\psi + g^{\mu\nu}
(i \bar\psi \gamma^{\lambda}\!\!\stackrel{\leftrightarrow}{\partial}\!\!_{\lambda}\psi
- m\bar\psi\psi),
\label{tfermionic}
\eeq
\beq
T^{\mu\nu}_{fp} = -\, e J^{(\mu}A^{\nu)} + e g^{\mu\nu}J^{\lambda}A_{\lambda}\,,
\eeq
and
\beq
T^{\mu\nu}_{ph} = F^{\mu\lambda}F^{\nu}_{\ \ \lambda} - \frac{1}{4} g^{\mu\nu}
F^{\lambda\rho}F_{\lambda\rho},
\label{tphoton}
\eeq
where the current is defined as
\beq
J^{\mu}(x) = \bar\psi (x) \gamma^{\mu} \psi (x)\,.\\
\label{vectorcurrent}
\eeq

The connected components of $TJJ$ can be obtained directly from the quantum average of $T_p$, defined
as the sum of the fermion contribution and its interaction part with the photon field,
\beq
T_p^{\mu\nu}\equiv T_{f}^{ \mu\nu} +T_{fp}^{ \mu\nu}.
\eeq

In the formalism of the background fields, the $TJJ$ correlator then can be extracted from the defining functional integral
\beqa
\langle T_p^{\mu\nu}(z)\rangle_A &\equiv& \int D\psi D\bar{\psi} \,\,T^{\mu\nu}_p (z) \,\,e^{i \int d^4 x \mathcal{L} + \int J\cdot A(x) d^4 x}\nonumber \\
&=& \langle T^{\mu\nu}_p \, e^{i \int d^4 x \, J\cdot A(x)}\rangle
\eeqa
via two functional derivatives respect to the background field $A_\mu$ and generates the effective action
\beq
\Gamma^{\mu\nu\alpha\beta} (z; x, y) \equiv \frac{ \delta^2 \lag T_p^{\mu\nu} (z) \rag_A}
{\delta A_{\alpha}(x)\delta A_{\beta}(y)} \bigg\vert_{A=0} = \Gamma_{anom}^{\mu\nu\alpha\beta} + \tilde{ \Gamma}^{\mu\nu\alpha\beta}.
\label{sep}
\eeq
We have separated in (\ref{sep}) the pole contribution $\Gamma_{anom}$ from the rest of the amplitude ($\tilde{\Gamma}$), which does not contribute to the trace part. Notice that
 $\Gamma_{anom}$, derived from either the classical generating functional (\ref{Tanom}) given by Riegert's action or from
 the direct perturbative expansion of (\ref{sep}), should nevertheless coincide, for the pole term not to be a spurious artifact of the variational solution. In particular, a computation performed in QED shows that the pole term extracted from $T_{anom}$ via functional differentiation

\beqa
\Gamma_{anom}^{\mu\nu\alpha\beta}(p,q) &=& \int\,d^4x\,\int\,d^4y\, e^{ip\cdot x + i q\cdot y}\,\frac{\delta^2 T^{\mu\nu}_{anom}(0)}
{\delta A_{\alpha}(x) A_{\beta}(y)} = \frac{e^2}{18\pi^2} \frac{1}{k^2} \left(g^{\mu\nu}k^2 - k^{\mu}k^{\nu}\right)u^{\alpha\beta}(p,q)
\nonumber \\
\label{Gamanom}
\eeqa
with
\beq
u^{\alpha\beta}(p,q) \equiv (p\cdot q) \,  g^{\alpha\beta} - q^{\alpha} \, p^{\beta}\,,\\
\label{utensor}
\eeq
indeed coincides with the result of the perturbative expansion, as defined from the first term on the rhs of (\ref{sep}). Thus, the entire contribution to the anomaly is extracted form $T_{anom}$ as
\beq
g_{\mu\nu}T^{\mu\nu}_{anom} = c F_{\alpha\beta}F^{\alpha\beta} = -\frac{e^2}{24\pi^2} F_{\alpha\beta}F^{\alpha\beta}.
\eeq
As we have already mentioned, the full action (\ref{Tnonl}), varied several times with respect to the background metric $g_{\mu\nu}$ and/or
the background gauge fields $A_{\alpha}$ gives those parts of the correlators of higher order, such as
$\lag TTT...JJ\rag$ and $\lag TTT...\rag$, which contribute to the trace anomaly. In particular, the anomalous contributions of
the $T^nJJ$'s vertices are obtained by varying the local action both respect to the metric and to the gauge fields.

\section{The energy momentum tensor and the Ward identities}
Moving to the QCD case, we introduce the definition of the QCD energy-momentum tensor, which is given by
\bea
T_{\mu \nu} &=& -g_{\mu \nu} {\mathcal L}_{QCD}
-F_{\mu \rho}^a F_\nu^{a \rho} -{\frac{1} {\xi}}g_{\mu \nu}
\partial^\rho (A_\rho^a \partial^\sigma A_\sigma^a) +{\frac{1}{\xi}}(A_\nu^a \partial_\mu(\partial^\sigma A^a_\sigma)
  +A_\mu^a \partial_\nu(\partial^\sigma A_\sigma^a))
\nonumber\\
&+& {\frac{i}{4}} \Big[
  \overline \psi \gamma_\mu (\overrightarrow \partial_\nu
-i g T^a A_\nu^a)\psi  -\overline \psi (\overleftarrow \partial_\nu
+i g T^a A_\nu^a)\gamma_\mu\psi
 +\overline \psi \gamma_\nu (\overrightarrow \partial_\mu
-i g T^a A_\mu^a)\psi
\nonumber\\
&-& \overline \psi (\overleftarrow \partial_\mu
+i g T^a A_\mu^a)\gamma_\nu\psi \Big] +\partial_\mu \overline
\omega^a (\partial_\nu \omega^a -g f^{abc} A_\nu^c \omega^b)
+\partial_\nu \overline \omega^a (\partial_\mu \omega^a -g f^{abc}
A_\mu^c \omega^b),
\label{EMT}
\eea
where $F_{\mu\nu}^a$ is the non-abelian field strength of the gauge field $A$
\bea
F_{\mu\nu}^a = \partial_{\mu}A_{\nu}^a - \partial_{\nu}A_{\mu}^a + g f^{abc}A^b_{\mu}A^c_{\nu}
\eea
and we have denoted with $\omega^a$ the Faddeev-Popov ghosts and with $\overline{\omega}^a$ the antighosts, while $\xi$ is the gauge-fixing parameter. The $T^a$'s are the gauge group generators in the fermion representation and $f^{abc}$ are the antisymmetric structure constants. For later use,
it is convenient to isolate the gauge-fixing and ghost dependent contributions from the entire tensor
\bea
T^{g.f.}_{\mu\nu} &=& {1 \over \xi}\left[A_\nu^a \partial_\mu(\partial \cdot A^a) +A_\mu^a \partial_\nu(\partial \cdot A^a)\right] -{1 \over \xi}g_{\mu \nu}
\left[- \frac{1}{2} (\partial \cdot A)^2 + \partial^\rho (A_\rho^a \partial \cdot A^a)\right], \\
T^{gh}_{\mu\nu} &=& \partial_{\mu}\bar\omega^a D^{ab}_{\nu}\omega^b + \partial_{\nu}\bar\omega^a D^{ab}_{\mu}\omega^b - g_{\mu\nu} \partial^{\rho}\bar\omega^a D^{ab}_{\rho}\omega^b.
\label{gfghost}
\eea

The coupling of QCD to gravity in the weak gravitational field limit is given by the interaction Lagrangian
\bea
\mathcal L_{int} = -\frac{1}{2} \kappa\, h^{\mu\nu} T_{\mu \nu}.
\eea
Notice that $T_{\mu \nu}$ as defined in Eq.~(\ref{EMT}) is symmetric, while traceless for a massless theory. The symmetric expression can be easily found as suggested in \cite{Freedman:1974gs}, by coupling the theory to gravity and then defining it via a functional derivative with respect to the metric, recovering (\ref{EMT}) in the flat spacetime case.

The conservation equation of the energy-momentum tensor takes the following form off-shell \cite{Nielsen:1975ph,Caracciolo:1989pt}
\bea
\partial^{\mu}T_{\mu\nu} &=& -\frac{\delta S}{\delta \psi} \partial_{\nu}\psi - \partial_{\nu}\bar \psi \frac{\delta S}{\delta \bar \psi} + \frac{1}{2}\partial^{\mu}\left( \frac{\delta S}{\delta \psi}\sigma_{\mu\nu}\psi - \bar \psi \sigma_{\mu\nu}\frac{\delta S}{\delta \bar \psi} \right) - \partial_{\nu}A_{\mu}^a \frac{\delta S}{\delta A_{\mu}^a} \nn\\
&+& \partial_{\mu}\left( A_{\nu}^a \frac{\delta S}{\delta A_{\mu}^a} \right) - \frac{\delta S}{\delta \omega^a} \partial_{\nu}\omega^a - \partial_{\nu}\bar \omega^a\frac{\delta S}{\delta \bar \omega^a} \,  \label{EMTdivergence}
\eea
where $\sigma_{\mu\nu}= \frac{1}{4}[\gamma_{\mu},\gamma_{\nu}]$. It is indeed conserved by using the equations of motion of the ghost, antighost and fermion/antifermion fields. The off-shell relation is particularly useful, since it can be inserted into the functional integral in order to derive some of the Ward identities satisfied by the correlator. In fact, the implications of the conservation of the energy-momentum tensor on the Green's functions can be exploited through the generating
functional, obviously defined as
\bea
&& Z[J,\eta,\bar\eta,\chi,\bar\chi,h] =
\int \mathcal D A \, \mathcal D \psi \, \mathcal D \bar\psi \,
\mathcal D \omega \, \mathcal D \bar \omega \, \exp\bigg\{ i \int d^4
x \left( \mathcal{L} + J_{\mu}A^{\mu} \right. \nn \\
&& \hspace{6cm} \left.  + \bar \eta \psi +
\bar \psi \eta + \bar \chi \omega + \bar \omega \chi +
h_{\mu\nu}T^{\mu\nu}\right) \bigg\},
\eea
where $\mathcal{L}$ is the standard QCD action and we have added the coupling of the
energy-momentum tensor of the theory to the background gravitational field $h_{\mu\nu}$, which is the typical expression needed in the study of QCD coupled to gravity with a linear deviation from the flat metric. We have denoted with $J,\eta,\bar\eta,\chi,\bar\chi$ the
sources of the gauge field $A$ ($J$), the source of the fermion and antifermion fields ($\bar{\eta}, \eta$) and of the ghost and antighost fields ($\bar{\chi}$, $\chi$) respectively. The generating functional $W$ of the connected Green's functions is, as usual, denoted by \bea \exp i \,
W[J,\eta,\bar\eta,\chi,\bar\chi,h] =
\frac{Z[J,\eta,\bar\eta,\chi,\bar\chi,h]}{Z[0,0,0,0,0,0]}
\eea
(normalized to the vacuum functional) and the effective action, defined as the generating functional $\Gamma$ of the 1-particle irreducible and truncated amplitudes. This is obviously obtained from $W$ by a Legendre transformation respect to all the sources, except, in our case, $h_{\mu\nu}$, which is taken as a background external field
\bea
\Gamma[A_c,\bar \psi_c, \psi_c,
\bar \omega_c, \omega_c, h] = W[J,\eta,\bar\eta,\chi,\bar\chi,h] -
\int d^4 x \left( J_{\mu}A^{\mu}_c + \bar \eta \psi_c + \bar \psi_c \eta
+ \bar \chi \omega_c + \bar \omega_c \chi \right).
\label{1PIfunctional}
\eea
The source fields are eliminated from the right hand side of Eq.~(\ref{1PIfunctional}) inverting the relations
\bea
A^{\mu}_c = \frac{\delta W }{\delta J_{\mu}}, \qquad
\psi_c = \frac{\delta W }{\delta \bar \eta}, \qquad \bar \psi_c =
\frac{\delta W }{\delta \eta}, \qquad \omega_c = \frac{\delta W
}{\delta \bar \chi}, \qquad \bar \omega_c = \frac{\delta W }{\delta
\chi} \label{Legendre1}
\eea
so that the functional derivatives of the effective action $\Gamma$ with respect to its independent variables are
\bea
\frac{\delta\Gamma}{\delta A^{\mu}_c} = - J_{\mu}, \qquad
\frac{\delta\Gamma}{\delta \psi_c} = - \bar \eta, \qquad
\frac{\delta\Gamma}{\delta \bar \psi_c} = - \eta, \qquad
\frac{\delta\Gamma}{\delta \omega_c} = - \bar \chi, \qquad
\frac{\delta\Gamma}{\delta \bar \omega_c} = - \chi,
 \label{Legendre2}
\eea
and for the source $h_{\mu\nu}$ we have instead
\bea
\frac{\delta\Gamma}{\delta h_{\mu\nu}} = \frac{\delta W}{\delta h_{\mu\nu}}. \label{Legendre3}
\eea
The conservation of the energy-momentum tensor summarized in Eq.~(\ref{EMTdivergence}) in terms of classical fields, can be re-expressed
in a functional form by a differentiation of $W$ with respect to $h_{\mu\nu}$ and the use of Eq.~(\ref{EMTdivergence})
under the functional integral. We obtain
\bea
\partial_{\mu} \frac{\delta W}{\delta h_{\mu\nu}} &=&  \bar \eta \, \partial_{\nu} \frac{\delta W}{\delta \bar \eta} +
\partial_{\nu} \frac{\delta W}{\delta \eta} \eta - \frac{1}{2}\partial^{\mu}\left(\bar \eta \sigma_{\mu\nu} \frac{\delta W}{\delta \bar \eta}-
 \frac{\delta W}{\delta \eta}\sigma_{\mu\nu}\eta  \right) + \partial_{\nu} \frac{\delta W}{\delta J_{\mu}} J_{\mu}
 - \partial_{\mu} \left( \frac{\delta W}{\delta J_{\mu} }J_{\nu} \right) \nn \\
 &+& \bar \chi \partial_{\nu}\frac{\delta W}{\delta \bar \chi} + \partial_{\nu}\frac{\delta W}{\delta \chi} \chi \,,
\label{divW}
\eea
and finally, for the one particle irreducible generating functional, this gives
\bea
\partial_{\mu} \frac{\delta \Gamma}{\delta h_{\mu\nu}} &=& - \frac{\delta \Gamma}{\delta \psi_c}\partial^{\nu}\psi_c
- \partial^{\nu}\bar\psi_c \frac{\delta \Gamma}{\delta \bar \psi_c} + \frac{1}{2}\partial_{\mu}\left( \frac{\delta \Gamma}{\delta \psi_c} \sigma^{\mu\nu} \psi_c
- \bar \psi_c \sigma^{\mu\nu}\frac{\delta \Gamma}{\delta \bar \psi_c}  \right) - \partial^{\nu}A^{\mu}_c\frac{\delta \Gamma}{\delta A^{\mu}_c}
+\partial^{\mu}\left(A^{\nu}_c\frac{\delta \Gamma}{\delta A^{\mu}_c} \right) \nn \\
&-& \frac{\delta \Gamma}{\delta \omega_c}\partial^{\nu}\omega_c - \partial^{\nu}\bar \omega_c \frac{\delta \Gamma}{\delta \bar \omega_c}
\label{FuncWI},
\eea
obtained from Eq.~(\ref{divW}) with the help of Eqs.~(\ref{Legendre1}, (\ref{Legendre2}), (\ref{Legendre3}). We summarize below the relevant Ward identities that can be used in order to fix the expression of the correlator.

\begin{itemize}
\item{\bf Single derivative general Ward identity}\\
The Ward identities describing the conservation of the energy-momentum tensor for the one-particle irreducible Green's functions with
an insertion of $T_{\mu\nu}$ can be obtained from the functional equation (\ref{FuncWI}) by taking functional derivatives with respect
to the classical fields. For example, the Ward identity for the graviton - gluon gluon vertex is obtained by differentiating Eq.~(\ref{FuncWI}) with respect to
$A_{c \, \alpha}^a(x_1)$ and $A_{c \, \beta}^b(x_2)$ and then setting all the external fields to zero
\bea
\partial^{\mu}\langle T_{\mu\nu}(x) A_\alpha^a (x_1)A_{\beta}^b (x_2)\rangle_{trunc}  &=& - \partial_{\nu}\delta^4(x_1-x) D^{-1}_{\alpha\beta}(x_2,x)
- \partial_{\nu}\delta^4(x_2-x) D^{-1}_{\alpha\beta}(x_1,x) \nn \\
&+& \partial^{\mu}\left( g_{\alpha \nu} \delta^4(x_1-x) D^{-1}_{\beta\mu}(x_2,x) + g_{\beta \nu} \delta^4(x_2-x) D^{-1}_{\alpha\mu}(x_1,x)\right) \label{WIcoord} \nonumber \\
\eea
where $D^{-1}_{\alpha\beta}(x_1,x_2)$ is the inverse gluon propagator defined as
\bea
D^{-1}_{\alpha\beta}(x_1,x_2) = \langle A_{\alpha}(x_1) A_{\beta}(x_2) \rangle_{trunc}  =  \frac{\delta^2 \Gamma}{\delta A^{\alpha}_c(x_1) \delta A^{\beta}_c(x_2) }
\eea
and where we have omitted, for simplicity, both the colour indices and the symbol of the $T$-product.
The first Ward identity (\ref{WIcoord}) becomes
\bea
k^{\mu}\langle T_{\mu\nu}(k) A_{\alpha}(p) A_{\beta}(q) \rangle_{trunc} =
q_{\mu} D^{-1}_{\alpha\mu}(p) g_{\beta\nu} + p_{\mu} D^{-1}_{\beta\mu}(q) g_{\alpha\nu}  - q_{\nu} D^{-1}_{\alpha\beta}(p) - p_{\nu} D^{-1}_{\alpha\beta}(q) \, .
\label{WImom}
\eea

\item{\bf Trace Ward identity at zero momentum transfer}

It is possible to extract a Ward identity for the trace of the energy-momentum tensor for the same correlation function using just Eq.~(\ref{WImom}).
In fact, differentiating it with respect to $p_{\mu}$ (or $q_{\mu}$) and then evaluating the resulting expression at zero momentum transfer ($p = - q$) we obtain the Ward identity in $d$ spacetime dimensions
\bea
\langle T^{\mu}_{\mu}(0) A_{\alpha}(p) A_{\beta}(-p) \rangle_{trunc} = \left(2 -d + p\cdot \frac{\partial}{\partial p} \right) D^{-1}_{\alpha\beta}(p)
\label{WItrace}
\eea
where the number $2$ counts the number of external gluon lines. For $d = 4$ and using the transversality of the one-particle irreducible self-energy, namely
\bea
D^{-1}_{\alpha\beta}(p) = (p^2 g_{\alpha \beta} - p_{\alpha} q_{\beta}) \Pi (p^2),
\eea
the Ward identity in Eq.~(\ref{WItrace}) simplifies to
\bea
\langle T^{\mu}_{\mu}(0) A_{\alpha}(p) A_{\beta}(-p) \rangle_{trunc} = 2 p^2(p^2 g_{\alpha \beta} - p_{\alpha} q_{\beta}) \frac{d \Pi}{d p^2}(p^2).
\eea
The trace Ward identity in Eq.~(\ref{WItrace}) at  zero momentum transfer  can also be explicitly related to the $\b$ function and the anomalous dimensions of the renormalized theory. These enter through the renormalization group equation for the two-point function of the gluon.
Defining the beta function and the anomalous dimensions  as
\bea
\beta(g) = \mu \frac{\partial g}{\partial \mu}, \qquad \gamma(g) = \mu \frac{\partial}{\partial \mu}\log\sqrt{Z_A} , \qquad m\, \gamma_m(g) = \mu \frac{\partial m}{\partial \mu}
\eea
and denoting with $Z_A$ the wave function renormalization constant of the gluon field, with $g$  the renormalized coupling, and with $m$ the renormalized mass, the trace Ward identity can be related to these functions by the relation
\bea
\langle T^{\mu}_{\mu}(0) A_{\alpha}(p) A_{\beta}(-p) \rangle_{trunc} = \left[ \beta(g)\frac{\partial}{\partial g} - 2 \gamma(g) + m (\gamma_m(g) -1) \frac{\partial}{\partial m}   \right] D^{-1}_{\alpha\beta}(p).
\label{traceJJ}
\eea

\item{\bf Two-derivatives Ward identity via BRST symmetry}
\end{itemize}
We can exploit the BRST symmetry of the gauge-fixed lagrangian in order to derive some generalized Ward (Slavnov-Taylor) identities. We start by computing the BRST variation of the energy-momentum tensor, which is given by
\bea
\delta A_{\mu}^a &=& \lambda D_{\mu}^{ab} \omega^b, \label{brsA} \\
\delta \omega^a &=& - \frac{1}{2} g \lambda f^{abc}\omega^b \omega^c, \label{brsO} \\
\delta \bar \omega^a &=& -\frac{1}{\xi}(\partial^{\mu} A_{\mu}^a)\lambda, \label{brsOb} \\
\delta \psi &=&  i g \lambda \omega^a t^a \psi, \\
\delta \bar \psi &=& - i g \bar \psi t^a \lambda \omega^a,
\eea
where $\lambda$ is an infinitesimal Grassmann parameter.\\
A careful analysis of the energy-momentum tensor presented in Eq.~(\ref{EMT}) shows that the fermionic and the gauge part are gauge invariant and therefore invariant also under BRST. Instead the gauge-fixing and the ghost contributions must be studied in more detail. Using the transformation equations (\ref{brsA}) and (\ref{brsOb}) in (\ref{gfghost}) one can prove the two identities
\bea
\lambda \, T^{g.f.}_{\mu\nu} &=& - A^{a}_{\nu} \partial_{\mu} \delta \bar\omega^a - A^{a}_{\mu} \partial_{\nu} \delta \bar\omega^a + g_{\mu\nu} \left[\frac{1}{2}\partial\cdot A^a \delta \bar\omega^a + A_{\rho}^a\partial^{\rho} \delta\bar\omega^a \right], \label{BRS_Tgf}\\
\lambda \, T^{gh}_{\mu\nu} &=& - \partial_{\mu}\bar \omega^a \delta A^a_{\nu} - \partial_{\nu}\bar \omega^a \delta A^a_{\mu} + g_{\mu\nu} \partial^{\rho}\bar\omega^a \delta A_{\rho}^a,
\label{BRS_Tgh}
\eea
which show that the ghost and the gauge-fixing parts of the energy-momentum tensor (times the anticommuting factor $\lambda$) can be written as an appropriate BRST variation of ghost/antighost and gauge contributions.
Their sum, instead, can be expressed as the BRST variation of a certain operator plus an extra term which vanishes when using the ghost equations of motion
\bea
\lambda \left( T^{g.f.}_{\mu\nu} + T^{gh}_{\mu\nu} \right) = \delta \left[ - \partial_{\mu}\bar \omega^a A_{\nu}^a - \partial_{\nu}\bar \omega^a A_{\mu}^a + g_{\mu\nu} \left( A_{\rho}^a \partial_{\rho} \bar \omega^a + \frac{1}{2} \partial\cdot A^a \omega^a \right) \right] + g_{\mu\nu} \frac{1}{2} \lambda \bar\omega^a \partial^{\rho} D^{ab}_{\rho}\omega^b,
\eea
which shows explicitly the structure of the gauge-variant terms in the energy-momentum tensor. Using the nilpotency of the
BRST operator ($\delta^2=0$), the BRST variation of $T_{\mu\nu}$ is given by
\bea
\delta T_{\mu\nu} = \delta (T_{\mu\nu}^{g.f.} + T_{\mu\nu}^{gh}) = \frac{\lambda}{\xi}\left[ A_{\mu}^a \partial_{\nu} \partial^{\rho} D^{ab}_{\rho}\omega^b + A_{\nu}^a \partial_{\mu} \partial^{\rho} D^{ab}_{\rho}\omega^b - g_{\mu\nu}\partial^{\sigma}(A_{\sigma}^a \partial^{\rho}D^{ab}_{\rho}\omega^b ) \right],
\eea
where it is straightforward to recognize the equation of motion of the ghost field on its right-hand side.
Using this last relation, we are able to derive some constraints on the Green's functions involving insertions of the energy-momentum tensor. In particular, we are interested in some identities satisfied by the correlator $\langle T_{\mu\nu} A_{\alpha}^a A_{\beta}^b \rangle$ in order to define it unambiguously. For this purpose, it is convenient to choose an appropriate  Green's function, in our case this is given by $\langle T_{\mu\nu} \partial^{\alpha} A_{\alpha}^a \bar \omega^b \rangle$, and then exploit its BRST invariance to obtain

\bea
\delta \langle T_{\mu\nu} \partial^{\alpha} A_{\alpha}^a \bar \omega^b \rangle = \langle \delta T_{\mu\nu} \partial^{\alpha} A_{\alpha}^a \bar \omega^b \rangle + \lambda \langle T_{\mu\nu} \partial^{\alpha} D^{ac}_{\alpha}\omega^c \bar \omega^b \rangle - \frac{\lambda}{\xi} \langle T_{\mu\nu} \partial^{\alpha} A_{\alpha}^a \partial^{\beta} A_{\beta}^b \rangle = 0,
\label{BRSTward}
\eea
where the first two correlators, built with operators proportional to the equations of motion, contribute only with disconnected amplitudes, that are not part of the one-particle irreducible vertex function.
 From Eq.~(\ref{BRSTward}) we obtain the identity
 \bea
\partial_{x_1}^{\alpha}\partial_{x_2}^{\beta}\langle T_{\mu\nu}(x) A_{\alpha}^a(x_1)A_{\beta}^b(x_2)\rangle_{trunc} = 0,
\eea
which in momentum space becomes
\bea
p^{\alpha} q^{\beta} \langle T_{\mu\nu}(k) A_{\alpha}^a(p)A_{\beta}^b(q)\rangle_{trunc} = 0. \label{brstwi}
\label{pract1}
\eea
A subtlety in these types of derivations concerns the role played by the commutators, which are generated because of the T-product and can be ignored only if they vanish. In general, in fact, the derivatives
are naively taken out of the correlator, in order to arrive at Eq.~ (\ref{pract1}) and this can generate an error. In this case, due to the presence of an energy momentum tensor, the evaluation of these terms is rather involved. For this reason one needs to perform an explicit check of (\ref{pract1}) to ensure the consistency of the formal result in a suitable regularization scheme.
As we are going to show in the next sections, these three Ward identities turn out to be satisfied in dimensional regularization.

\section{The perturbative expansion}
\begin{figure}[t]
\begin{center}
\includegraphics[scale=0.9]{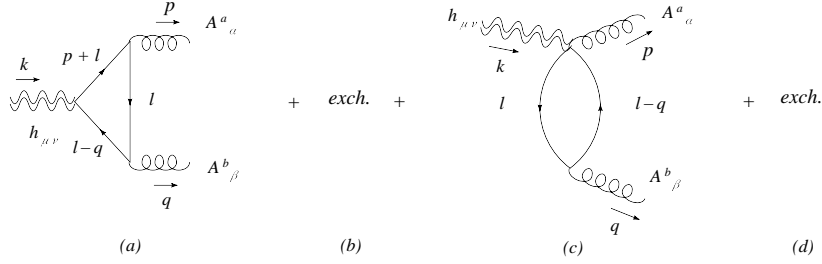}
\caption{\small The fermionic contributions  with a graviton $h_{\mu\nu}$ in the initial state and two gluons $A^a_\a, A^b_\b$ in the final state. }
\label{fermloop}
\end{center}
\end{figure}

\begin{figure}[t]
\begin{center}
\includegraphics[scale=0.9]{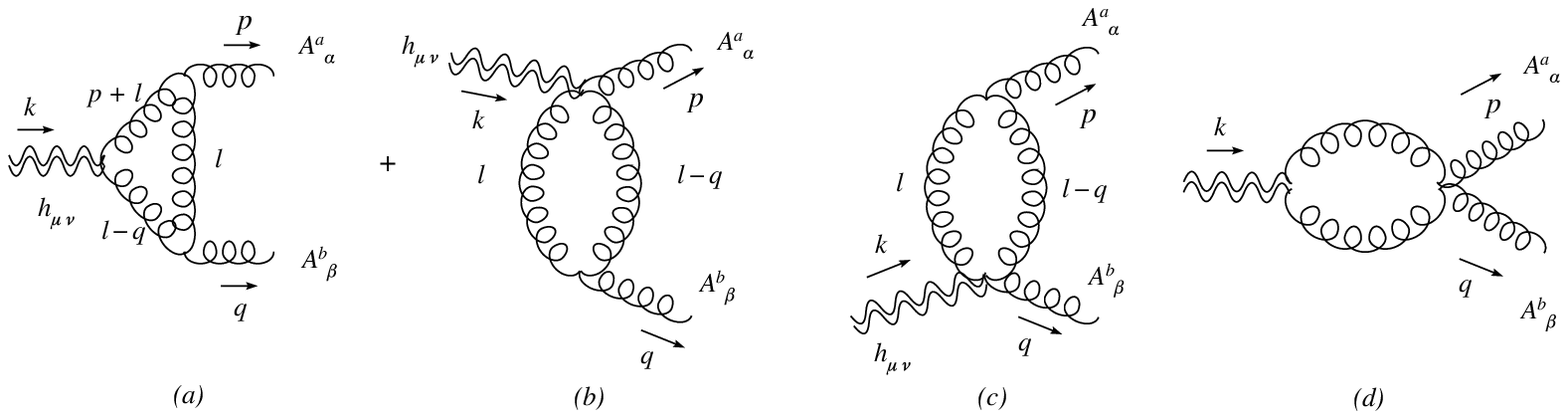}
\caption{\small The gauge contributions  with a graviton $h_{\mu\nu}$ in the initial state and two gluons $A^a_\a, A^b_\b$ in the final state. }
\label{gluonloop}
\end{center}
\end{figure}

\begin{figure}[t]
\begin{center}
\includegraphics[scale=0.9]{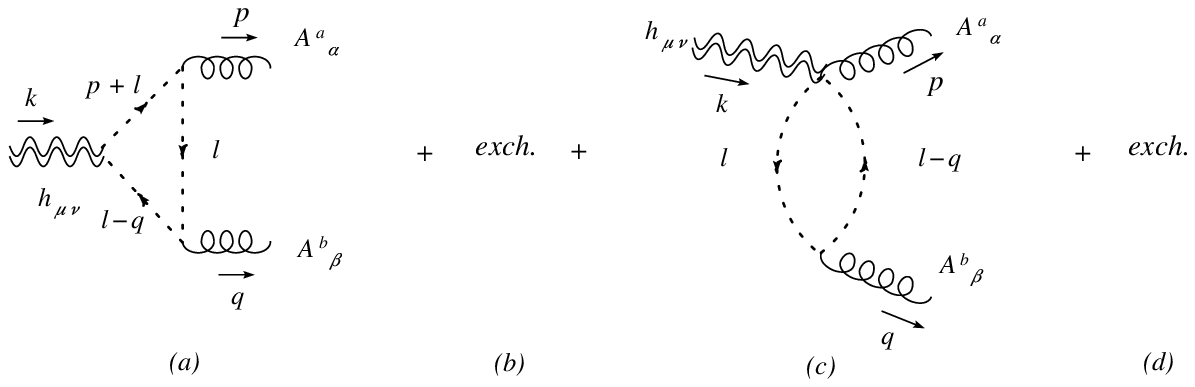}
\caption{\small The ghost contributions  with a graviton $h_{\mu\nu}$ in the initial state and two gluons $A^a_\a, A^b_\b$ in the final state. }
\label{ghostloop}
\end{center}
\end{figure}

The perturbative expansion is obtained by taking into account all the diagrams depicted in Figs.~\ref{fermloop}, \ref{gluonloop}, \ref{ghostloop}, where an incoming graviton appears in the initial state and two gluons with momenta $p$ and $q$ characterize the final state. The different contributions to the total amplitude are identified by the nature of the internal lines and are computed with the aid of the Feynman rules defined  in Appendix \ref{rules}. Each amplitude is denoted by $\Gamma$, with a superscript in square brackets indicating the figure of the corresponding diagram.

The contributions with  a massive fermion running in the loop are depicted in  Fig.~\ref{fermloop}; for the triangle in Fig.~\ref{fermloop}a we obtain
\bea
- i \frac{\kappa}{2}\, \Gamma^{{\bf[2a]} \,ab}_{\mu\nu\alpha\beta}(p,q) = - \frac{\kappa}{2} g^2 \,{\rm Tr}(T^b T^a) \int\frac{d^4 l}{(2\pi)^4} {\rm tr}\left\{V^{\prime}_{\mu\nu}(l-q,l+p)\frac{1}{\lsl-\qsl-m}\gamma_{\beta}\frac{1}{\lsl-m}\gamma_{\alpha}\frac{1}{\lsl+\psl-m}  \right\}
\eea
where the color factor is given by ${\rm Tr}(T^b T^a) = \frac{1}{2} \delta^{a b}$; the diagram in Fig.~\ref{fermloop}c contributes as
\bea
- i \frac{\kappa}{2} \, \Gamma^{{\bf[2c]} \, ab}_{\mu\nu\alpha\beta}(p,q) = - \frac{\kappa}{2} g^2 \,{\rm Tr}(T^a T^b) \int\frac{d^4 l}{(2\pi)^4} {\rm tr}\left\{W^{\prime}_{\mu\nu\alpha}\frac{1}{\lsl-\qsl-m}\gamma_{\beta}\frac{1}{\lsl-m} \right\},
\eea
with the vertices $V^{\prime}_{\mu\nu}(l-q,l+p)$ and $W^{\prime}_{\mu\nu\alpha}$ defined in Appendix \ref{rules}, Eqs.~(\ref{VGff}) and (\ref{WGffg}) respectively.
The remaining diagrams in Fig.~\ref{fermloop} are obtained by exchanging
$\alpha \leftrightarrow \beta$ and $p \leftrightarrow q$
\bea
- i \, \frac{\kappa}{2} \, \Gamma^{{\bf[2b]} \,ab}_{\mu\nu\alpha\beta}(p,q) &=&
- i \, \frac{\kappa}{2} \, \Gamma^{{\bf[2a]} \,ab}_{\mu\nu\alpha\beta}(p,q)\bigg|_{
\bmi[c]{.2\linewidth}
\footnotesize{$\alpha \leftrightarrow \beta$}
\vspace{-.2cm} \\
 \footnotesize {$p \leftrightarrow q$}
 \emi
}
\hspace{-2cm},
\\ \nn \\
 - i \, \frac{\kappa}{2} \, \Gamma^{{\bf[2d]} \,ab}_{\mu\nu\alpha\beta}(p,q) &=&
 - i \, \frac{\kappa}{2} \, \Gamma^{{\bf[2c]} \,ab}_{\mu\nu\alpha\beta}(p,q)\bigg|_{
 \bmi[c]{.2\linewidth}
 \footnotesize{$\alpha \leftrightarrow \beta$}
 \vspace{-.2cm} \\
  \footnotesize {$p \leftrightarrow q$}
  \emi
} \hspace{-2cm}.
\eea
Moving to the gauge sector we find the four contributions in Fig.~\ref{gluonloop}: the first one with a triangular topology is given by
\bea
- i \frac{\kappa}{2} \, \Gamma^{{\bf[3a]} \,ab}_{\mu\nu\alpha\beta}(p,q) = - \frac{\kappa}{2} g^2 f^{ade}f^{bde} \int\frac{d^4 l}{(2\pi)^4} \frac{1}{l^2\,(l+p)^2\,(l-q)^2} \bigg[V^{Ggg}_{\mu\nu\rho\sigma}(l-q,-l-p)\,\times  \nn \\
 V^3_{\tau\sigma\alpha}(-l,l+p,-p)\, \, V^3_{\rho\tau\beta}(-l+q,l,-q)\bigg],
\eea
where the color factor is $f^{ade}f^{bde} = C_A \, \delta^{a b}$. Those in Figs.~\ref{gluonloop}b and \ref{gluonloop}c, containing
gluon loops attached to the graviton vertex, are called  ``t-bubbles'' and can be obtained one from the other by the exchange of $\alpha \leftrightarrow \beta$ and $p \leftrightarrow q$. The first ``t-bubble'' is given by
\bea
- i \frac{\kappa}{2} \, \Gamma^{{\bf[3b]} \,ab}_{\mu\nu\alpha\beta}(p,q) = - \frac{1}{2}\frac{\kappa}{2} g^2 f^{ade}f^{bde} \int\frac{d^4 l}{(2\pi)^4} \frac{V^{Gggg}_{\mu\nu\rho\sigma\beta}(-l,l-p,-q)\,V^3_{\rho\alpha\sigma}(k,-p,-l+p)}{l^2\,(l-p)^2}
\eea
which is multiplied by an additional symmetry factor $\frac{1}{2}$. There is another similar contribution obtained from the previous one after exchanging $\alpha \leftrightarrow \beta$ and $p \leftrightarrow q$
\bea
- i \frac{\kappa}{2} \, \Gamma^{{\bf[3c]} \,ab}_{\mu\nu\alpha\beta}(p,q) =
- i \frac{\kappa}{2} \, \Gamma^{{\bf[3b]} \,ab}_{\mu\nu\alpha\beta}(p,q)\bigg|_{
 \bmi[c]{.25\linewidth}
 \footnotesize{$\alpha \leftrightarrow \beta$}
 \vspace{-.2cm} \\
  \footnotesize {$p \leftrightarrow q$}
  \emi
}\hspace{-3cm}.
\eea
The last diagram with gluons running in the loop is the one in Fig.~\ref{gluonloop}d which is given by
\bea
- i \frac{\kappa}{2} \, \Gamma^{{\bf[3d]} ab}_{\mu\nu\alpha\beta}(p,q) = \frac{1}{2}\frac{\kappa}{2} g^2 \int\frac{d^4 l}{(2\pi)^4} \frac{
V^{Ggg}_{\, \mu\nu\rho\sigma}(-l,l-p-q)\, \delta^{d f}\, \, V^{4\,abcd}_{\, \rho\alpha\sigma\beta} }{l^2\,(l-p-q)^2},
\label{gluond}
\eea
where $V^4$ is the four gluon vertex defined as
\bea
-i g^2 V^{4\,abcd}_{\mu\nu\rho\sigma} &=&
-i g^2\left[ f^{abe}f^{cde}(g_{\mu\rho}g_{\nu\sigma} - g_{\mu\sigma}g_{\nu\rho})
+ f^{ace}f^{bde}(g_{\mu\nu}g_{\rho\sigma} - g_{\mu\sigma}g_{\nu\rho})  \right. \nn  \\
&& \hspace{2cm} \left. + \, f^{ade}f^{bce}(g_{\mu\nu}g_{\rho\sigma} - g_{\mu\rho}g_{\nu\sigma}) \right]
\eea
and therefore
\bea
\delta^{d f}\,V^{4\,abcd}_{\rho\alpha\sigma\beta} &=& - C_A \delta^{a b} \tilde{V}^4_{\rho\alpha\sigma\beta} = - C_A \delta^{a b} \left( g_{\alpha\sigma}g_{\beta\rho} +  g_{\alpha\rho}g_{\beta\sigma} - 2 g_{\alpha\beta}g_{\sigma\rho} \right),
\eea
so that the amplitude in Eq.~(\ref{gluond}) becomes
\bea
- i \frac{\kappa}{2} \, \Gamma^{{\bf[3d]}\, ab}_{\mu\nu\alpha\beta}(p,q) = - \frac{1}{2}\frac{\kappa}{2} g^2 C_A \delta^{a b} \int\frac{d^4 l}{(2\pi)^4} \frac{V^{Ggg}_{\mu\nu\rho\sigma}(-l,l-p-q)\, \tilde{V}^4_{\rho\alpha\sigma\beta} }{l^2\,(l-p-q)^2}.
\eea
In the expression above we have explicitly isolated the color factor $C_A \delta^{a b}$ and the symmetry factor $\frac{1}{2}$.

Finally, the ghost contributions shown in Fig.~\ref{ghostloop} are  given by the sum of
\bea
- i \frac{\kappa}{2} \, \Gamma^{{\bf[4a]}\, ab}_{\mu\nu\alpha\beta}(p,q) = - \frac{\kappa}{2} g^2 f^{ade}f^{bde} \int\frac{d^4 l}{(2\pi)^4} \frac{C_{\mu\nu\rho\sigma}(l-q)^{\rho}(l+p)^{\sigma}l_{\alpha}(l-q)_{\beta} }{l^2\, (l+p)^2 \, (l-q)^2}
\label{ghosta}
\eea
for the triangle diagram in Fig.~\ref{ghostloop}a and
\bea
- i \frac{\kappa}{2} \, \Gamma^{{\bf[4b]} \, ab}_{\mu\nu\alpha\beta}(p,q) = \frac{\kappa}{2} g^2 f^{ade}f^{bde} \int\frac{d^4 l}{(2\pi)^4} \frac{C_{\mu\nu\alpha\sigma}l^{\sigma}(l-q)_{\beta} }{l^2 \, (l-q)^2}
\label{ghostb}
\eea
for the ``T-bubble" diagram shown in Fig.~\ref{ghostloop}c. The two exchanged diagrams are obtained from those in Eqs.~(\ref{ghosta}) and (\ref{ghostb}) with the usual replacement $\alpha \leftrightarrow \beta$ and $p \leftrightarrow q$.
\bea
- i \, \frac{\kappa}{2} \, \Gamma^{{\bf[4b]} \,ab}_{\mu\nu\alpha\beta}(p,q) &=&
- i \, \frac{\kappa}{2} \, \Gamma^{{\bf[4a]} \,ab}_{\mu\nu\alpha\beta}(p,q)\bigg|_{
\bmi[c]{.2\linewidth}
\footnotesize{$\alpha \leftrightarrow \beta$}
\vspace{-.2cm} \\
 \footnotesize {$p \leftrightarrow q$}
 \emi
}
\hspace{-2cm},
\\ \nn \\
 - i \, \frac{\kappa}{2} \, \Gamma^{{\bf[4d]} \,ab}_{\mu\nu\alpha\beta}(p,q) &=&
 - i \, \frac{\kappa}{2} \, \Gamma^{{\bf[4c]} \,ab}_{\mu\nu\alpha\beta}(p,q)\bigg|_{
 \bmi[c]{.2\linewidth}
 \footnotesize{$\alpha \leftrightarrow \beta$}
 \vspace{-.2cm} \\
  \footnotesize {$p \leftrightarrow q$}
  \emi
} \hspace{-2cm}.
\eea
Having identified the different sectors we obtain the total amplitude for quarks, denoted by a ``q'' subscript
\bea
\Gamma_{q, \, \mu\nu\a\b}^{ab}(p,q) =
\Gamma^{{\bf[2a]} \,ab}_{\mu\nu\alpha\beta}(p,q) +
\Gamma^{{\bf[2b]} \,ab}_{\mu\nu\alpha\beta}(p,q) +
\Gamma^{{\bf[2c]} \,ab}_{\mu\nu\alpha\beta}(p,q) +
\Gamma^{{\bf[2d]} \,ab}_{\mu\nu\alpha\beta}(p,q)
\label{Gammaq}
\eea
and the one for gluons and ghosts as
\bea
\Gamma_{g, \, \mu\nu\a\b}^{ab}(p,q) =
\sum_{j=3,4} \, \left[ \Gamma^{{\bf[ja]} \,ab}_{\mu\nu\alpha\beta}(p,q) +
\Gamma^{{\bf[jb]} \,ab}_{\mu\nu\alpha\beta}(p,q) +
\Gamma^{{\bf[jc]} \,ab}_{\mu\nu\alpha\beta}(p,q) +
\Gamma^{{\bf[jd]} \,ab}_{\mu\nu\alpha\beta}(p,q) \right].
\label{Gammag}
\eea
\section{The on-shell $\langle TAA \rangle$ correlator, pole terms and form factors}
We proceed with a classification of all the diagrams contributing to the on-shell vertex, starting from the gauge invariant subset of diagrams that involve fermion loops and then moving to the second set, the one relative to gluons and ghosts. The analysis follows rather closely the method presented in the case of QED in previous works \cite{Giannotti:2008cv, Armillis:2009pq}, with a classification of all the relevant tensor structures which can be generated using the 43 monomials built out of the 2 of the 3 external momenta of the triangle diagram and the metric tensor $g_{\mu\nu}$. In general, one can proceed
with the identification of a subset of these tensor structure which allow to formulate the final expression in a manageable form. The fermionic triangle diagrams, which define one of the two gauge invariant subsets of the entire correlator, can be given in a simplified form also for off mass-shell external momenta, in terms of 13 form factors as in \cite{Giannotti:2008cv, Armillis:2009pq} while the structure of the gluon contributions are more involved. Some drastic semplifications take place in the on-shell case, where only 3 form factors - both in the quark and fermion sectors - are necessary to describe the final result.

We write the whole amplitude $\Gamma^{\mu\nu\a\b}(p,q)$ as
\bea
\Gamma^{\mu\nu\a\b}(p,q) =
\Gamma_q^{\mu\nu\a\b}(p,q) + \Gamma_g^{\mu\nu\a\b}(p,q),
\eea
 referring respectively to the contributions with quarks $(\Gamma_q)$ and with gluons/ghosts $(\Gamma_g)$ in Eqs.~(\ref{Gammaq}) and (\ref{Gammag}). We have omitted the color indices for simplicity. The amplitude $\Gamma$ is expressed in terms of 3 tensor structures and 3 form factors renormalized in the $\overline{MS}$ scheme
\beq
\Gamma^{\mu\nu\alpha\beta}_{q/g}(p,q) =  \, \sum_{i=1}^{3} \Phi_{i\,q/g} (s,0, 0,m^2)\, \delta^{ab}\, \phi_i^{\mu\nu\alpha\beta}(p,q)\,.
\label{Gamt}
\eeq
One comment concerning the choice of this basis is in order. The 3 form factors are more easily identified in the fermion sector after performing the on-shell limit of the off-shell amplitude, where the 13 form factors introduced in \cite{Giannotti:2008cv, Armillis:2009pq} for QED simplify into the 3 tensor structures that will be given below. It is then observed that the  tensor structure of the gluon sector, originally expressed in terms of the 43 monomials of  \cite{Giannotti:2008cv, Armillis:2009pq}, can be arranged consistently in terms of these 3 reduced structures.

The tensor basis on which we expand the on-shell vertex is given by
\bea
  \phi_1^{\, \mu \nu \a \b} (p,q) &=&
 (s \, g^{\mu\nu} - k^{\mu}k^{\nu}) \, u^{\a \b} (p,q),
 \label{widetilde1}\\
\phi_2^{\, \mu \nu \a \b} (p,q) &=& - 2 \, u^{\a \b} (p,q) \left[ s \, g^{\mu \nu} + 2 (p^\mu \, p^\nu + q^\mu \, q^\nu )
- 4 \, (p^\mu \, q^\nu + q^\mu \, p^\nu) \right],
\label{widetilde2} \\
\phi^{\, \mu \nu \alpha \beta}_{3} (p,q) &=&
\big(p^{\mu} q^{\nu} + p^{\nu} q^{\mu}\big)g^{\alpha\beta}
+ \frac{s}{2} \left(g^{\alpha\nu} g^{\beta\mu} + g^{\alpha\mu} g^{\beta\nu}\right) \nn \\
&&  \hspace{1cm} - g^{\mu\nu} \left(\frac{s}{2} g^{\alpha \beta}- q^{\alpha} p^{\beta}\right)
-\left(g^{\beta\nu} p^{\mu}
+ g^{\beta\mu} p^{\nu}\right)q^{\alpha}
 - \big (g^{\alpha\nu} q^{\mu}
+ g^{\alpha\mu} q^{\nu }\big)p^{\beta},
\label{widetilde3}
\nn \\
\eea
where $u^{\a \b} (p,q)$ has been defined in Eq.~(\ref{utensor}).
The form factors $\Phi_i(s,s_1,s_2,m^2)$ have as entry variables, beside $s=(p+q)^2$, the virtualities of the two gluons $s_1=p^2$ and $s_2=q^2$.\\
In the on-shell case only 3 invariant amplitudes contribute, which for the quark loop amplitude are given by
\bea
\Phi_{1\, q} (s,\,0,\,0,\,m^2) &=&
- \frac{g^2 }{36 \pi^2  s} \,  + \, \frac{g^2  m^2}{6 \pi^2 s^2} \, - \, \frac{g^2\, m^2} {6 \pi^2 s}\mathcal C_0 (s, 0, 0, m^2)
\bigg[\frac{1}{2 \,  }-\frac{2 m^2}{ s}\bigg],  \\
\Phi_{2\, q} (s,\,0,\,0,\,m^2)  &=&
- \frac{g ^2}{288 \pi^2 s} - \frac{ g ^2 m^2}{24 \pi^2  s^2}
- \, \frac{g^2 \, m^2}{8 \pi^2  s^2} \mathcal D (s, 0, 0, m^2) \,
\nn \\
&& \hspace{2cm}- \, \frac{g^2 \, m^2}{12 \pi^2 s } \mathcal C_0(s, 0, 0, m^2 )\, \left[ \frac{1}{2} + \frac{m^2}{s}\right],  \\
\Phi_{3\,q} (s,\,0,\,0,\,m^2) &=& \frac{11  g ^2}{288  \pi^2 }  +   \frac{ g ^2 m^2}{8 \pi^2 s}
 + \, g^2\mathcal C_0 (s, 0,0,m^2) \,\left[ \frac{m^4}{4 \pi^2 s}+\frac{ m^2}{8 \pi^2 }\right] \nn \\
 && \hspace{2cm} +  \frac{5 \,g^2 \, m^2}{24 \pi^2  s}  \mathcal D (s, 0, 0, m^2) + \frac{g^2}{24 \pi^2} \mathcal B_0^{\overline{MS}}(s, m^2),
\label{masslesslimit}
\eea
where the on-shell scalar integrals $\mathcal D (s, 0, 0, m^2)$, $\mathcal C_0 (s, 0,0,m^2)$ and $B_0^{\overline{MS}}(s, m^2)$ are computed in Appendix \ref{scalars}. In the massless limit the amplitude $\Gamma_q^{\mu\nu\a\b}(p,q)$ takes a simpler expression and the previous form factors become
\bea
\Phi_{1\,q} (s, 0, 0, 0) &=& - \frac{g^2}{36 \pi^2  s}, \\
\Phi_{2\,q} (s, 0, 0, 0) &=& - \frac{g^2}{288 \pi^2 \, s}, \\
\Phi_{3\,q} (s, 0, 0, 0) &=& - \frac{g^2}{288 \pi^2} \, \left[ 12 L_s - 35\right],
\eea
\begin{figure}[t]
\begin{center}
\includegraphics[scale=1.0]{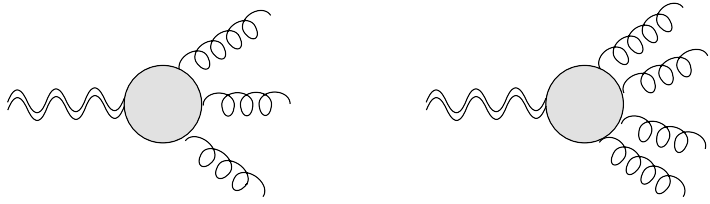}
\caption{\small Higher order contributions to the anomaly pole involved in the covariantization of the graviton/2-gluons amplitude. }
\label{additional}
\end{center}
\end{figure}

where
\bea
L_s \equiv \log \left( - \frac{s}{\mu^2} \right ) \qquad \qquad s<0.
\eea
In the gluon sector the computation of $\Gamma_g^{\mu\nu\a\b}(p,q)$ is performed analogously by using dimensional regularization with modified minimal subtraction ($\overline{MS}$) and we obtain for  on-shell gluons
\beq
\Gamma^{\mu\nu\alpha\beta}_{g}(p,q) =  \, \sum_{i=1}^{3} \Phi_{i\,g} (s,0,0)\,\delta^{ab}\, \phi_i^{\mu\nu\alpha\beta}(p,q)\, ,
\label{gl1}
\eeq
where the form factors obtained from the explicit computation are
\bea
\Phi_{1\,g}(s,0,0) &=& \frac{11 \, g^2}{72 \pi^2 \, s} \, C_A,\\
\Phi_{2\,g}(s,0,0) &=& \frac{g^2}{288 \pi^2 \, s} \, C_A, \\
\Phi_{3\,g}(s,0,0) &=& - g^2 \, C_A \bigg[ \frac{65}{288 \pi^2} + \frac{11}{48 \pi^2} \mathcal B_0^{\overline{MS}}(s,0) - \frac{1}{8 \pi^2}\mathcal B_0^{\overline{MS}}(0,0) +  \frac{s}{8 \pi^2}  \,\mathcal C_0(s,0,0,0) \bigg].
\label{gl2}
\eea
The renormalized scalar integrals can be found in Appendix \ref{scalars}.

The full on-shell vertex, which is the sum of the quark and pure gauge contributions, can be decomposed by using  the same three tensor structures $\phi_i^{\mu\nu\alpha\beta}$ appearing in the expansion of $\Gamma_q^{\mu\nu\a\b}(p,q)$ and $\Gamma_g^{\mu\nu\a\b}(p,q)$

\bea
\Gamma^{\mu\nu\alpha\beta}(p,q) =  \Gamma^{\mu\nu\alpha\beta}_g(p,q) + \Gamma^{\mu\nu\alpha\beta}_q(p,q) =  \sum_{i=1}^{3} \Phi_{i} (s,0, 0)\, \delta^{ab}\, \phi_i^{\mu\nu\alpha\beta}(p,q)\,,
\eea
with form factors defined as

\bea
\Phi_i(s,0,0) = \Phi_{i,\,g}(s,0,0) + \sum_{j=1}^{n_f}\Phi_{i, \,q}(s,0,0,m_j^2),
\eea

where the sum runs over the $n_f$ quark flavors. In particular we have
\bea
\Phi_{1}(s,0,0) &=& - \frac{g^2}{72 \pi^2 \,s}\left(2 n_f - 11 C_A\right) + \frac{g^2}{6 \pi^2}\sum_{i=1}^{n_f} m_i^2 \, \bigg\{ \frac{1}{s^2} \, - \, \frac{1} {2 s}\mathcal C_0 (s, 0, 0, m_i^2)
\bigg[1-\frac{4 m_i^2}{ s}\bigg] \bigg\}, \,
\label{Phi1}\\
\Phi_{2}(s,0,0) &=& - \frac{g^2}{288 \pi^2 \,s}\left(n_f - C_A\right) \nn \\
&&- \frac{g^2}{24 \pi^2} \sum_{i=1}^{n_f} m_i^2 \, \bigg\{ \frac{1}{s^2}
+ \frac{ 3}{ s^2} \mathcal D (s, 0, 0, m_i^2)
+ \frac{ 1}{s } \mathcal C_0(s, 0, 0, m_i^2 )\, \left[ 1 + \frac{2 m_i^2}{s}\right]\bigg\},
\label{Phi2} \\
\Phi_{3}(s,0,0) &=& \frac{g^2}{288 \pi^2}\left(11 n_f - 65 C_A\right) - \frac{g^2 \, C_A}{8 \pi^2} \bigg[ \frac{11}{6} \mathcal B_0^{\overline{MS}}(s,0) - \mathcal B_0^{\overline{MS}}(0,0) +  s  \,\mathcal C_0(s,0,0,0) \bigg] \nn \\
&& + \, \frac{g^2}{8 \pi^2} \sum_{i=1}^{n_f}\bigg\{  \frac{1}{3}\mathcal B_0^{\overline{MS}}(s, m_i^2) + m_i^2 \, \bigg[
\frac{1}{s}
 + \frac{5}{3 s}  \mathcal D (s, 0, 0, m_i^2) + \mathcal C_0 (s, 0,0,m_i^2) \,\left[1 + \frac{2 m_i^2}{s}\right]
\bigg]\bigg\} ,\nn \\
\label{Phi3}
\eea
with $C_A = N_C$ and the scalar integrals defined in Appendix \ref{scalars}.
Notice the appearance in the total amplitude of the $1/s$ pole in $\Phi_1$, which is present both in the quark and in the gluon sectors, and which saturates the contribution to the trace anomaly in the massless limit. In this case the entire trace anomaly is just proportional to this component, which becomes
\beq
\Phi_{1}(s,0,0) = - \frac{g^2}{72 \pi^2 \,s}\left(2 n_f - 11 C_A\right).
\label{polepole}
\eeq

The correlator $\Gamma^{\mu\nu\alpha\beta}(p,q)$, computed using dimensional regularization, satisfies all the Ward identities defined in the previous sections. Notice that the two-derivatives Ward identity introduced in Eq.~(\ref{brstwi})
\bea
p_{\a}  q_{\b} \, \Gamma^{\mu\nu\a\b}(p,q) = 0,
\eea
derived from the BRST symmetry of the QCD Lagrangian, is straightforwardly satisfied by the on-shell amplitude. This is easily seen from the tensor decomposition introduced in Eq.~(\ref{Gamt}) because all the tensors fulfill the condition
\bea
p_{\a}  q_{\b} \,  \phi_1^{\, \mu \nu \a \b} (p,q)  = 0.
\eea
Furthermore, we have checked at one-loop order the validity of the single derivative Ward identity given in Eq.~(\ref{WImom}) and describing the conservation of the energy-momentum tensor. Using the transversality of the two-point gluon function Eq.~(\ref{WImom}) this gives
\bea
k_{\mu} \, \Gamma^{\mu\nu\a\b}(p,q) =
\left( q^\nu \, p^\a \, p^\b - q^\nu \, g^{\a \b} \, p^2 + g^{\nu \b} \, q^\a \, p^2 - g^{\nu \b} \, p^\a \, p \cdot q \right) \, \Pi (p^2) \nn \\
+ \left( p^\nu \, q^\a \, q^\b - p^\nu \, g^{\a \b} \, q^2 + g^{\nu \a} \, p^\b \, q^2 - g^{\nu \a} \, q^\b \, p \cdot q \right) \, \Pi (q^2),
\eea
where the renormalized gluon self energies are defined as
\bea
\Pi (p^2) &=& \frac{g^2 \, C_A \, \d^{a b} \, }{144 \, \pi^2} \, \left( 15 \, \mathcal B_0^{\overline{MS}} (p^2, 0) -2 \right)  \nn \\
&+& \frac{g^2 \, \d^{ab}}{72 \, \pi^2 p^2}\sum_{i=1}^{n_f} \, \left[ 6 \, \mathcal A_0^{\overline{MS}} \,  (m_i^2) + p^2 - 6 \, m_i^2
- 3 \, \mathcal B_0^{\overline{MS}} (p^2, m_i^2) \left( 2 \, m_i^2 + p^2 \right)    \right]\, .
\label{Pigluon}
\eea
The QCD $\beta$ function can be related to the residue of the pole and can be easily computed starting from the amplitude  $\Gamma^{\mu\nu\alpha\beta}(p,q)$ for on-shell external lines and in the conformal limit
\bea
g_{\mu\nu}\,  \Gamma^{\mu\nu\alpha\beta}(p,q) =  3 \, s \, \Phi_1 (s; 0,0,0) \, u^{\a \b}(p,q) = - 2 \, \frac{\b (g)}{g}\, u^{\a \b}(p,q),
\eea
with the QCD $\b$ function given by
\bea
\b (g)= \frac{g^3}{16 \pi^2} \left (  - \frac{11}{3} \, C_A + \frac{2}{3} \, n_f \right) .
\eea
As we have already mentioned, after contracting the metric tensor $g_{\mu\nu}$ with the whole amplitude $\Gamma $, only the  tensor structure $\phi^{\mu\nu\a\b}_1(p,q)$ contributes to the anomaly,  being the remaining ones traceless, with a contribution entirely given by $\Phi_1|_{m=0}$  in Eq.~(\ref{Phi1}), i.e. Eq. (\ref{polepole}). In the massive fermion case, the anomalous contribution are corrected by terms proportional to the fermion mass $m$ and represent an explicit breaking of scale invariance. From a direct computation we can also extract quite straightforwardly the effective action, which is given by
\bea
S_{pole} &=&	- \frac{c}{6}\, \int d^4 x \, d^4 y \,R^{(1)}(x)\, \square^{-1}(x,y) \,  F^a_{\alpha \beta} \,  F^{a \, \alpha \beta} \nn\\
&=& \frac{1}{3} \, \frac{g^3}{16 \pi^2} \left (  - \frac{11}{3} \, C_A + \frac{2}{3} \, n_f \right)  \, \int d^4 x \, d^4 y \,R^{(1)}(x)\, \square^{-1}(x,y) \, F_{\alpha \beta}F^{\alpha \beta}
\eea
and is in agreement with Eq. (\ref{SSimple}), derived from the nonlocal gravitational action.
Here $R^{(1)}$ denotes the linearized expression of the Ricci scalar
\beq
 R^{(1)}_x\equiv \partial^x_\mu\, \partial^x_\nu \, h^{\mu\nu} - \square \,  h, \qquad h=\eta_{\mu\nu} \, h^{\mu\nu}
 \eeq
and the constant $c$ is related to the non-abelian $\beta$ function as
\beq
c= - 2 \, \, \frac{\beta (g)}{g}.
\eeq
Notice that the contribution coming from $TJJ$ generates the abelian part of the non-abelian field strength, while extra contributions (proportional to extra factors of $g$ and $g^2$) are expected from the $TJJJ$ and $TJJJJ$ diagrams (see Fig.~\ref{additional}). This situation is analogous to that of the gauge anomaly, where one needs to render gauge covariant the anomalous amplitude given by the triangle diagram. In that case the gauge covariant expression is obtained by adding to the $AVV$ vertex also the $AVVV$ and $AVVVV$ diagrams, with 3 and 4 external gauge lines, respectively.


\section{Comments}
The appearance of massless degrees of freedom in the effective action describing the coupling of gravity to the gauge fields
is rather intriguing, and is an aspect that will require further analysis. 

The nonlocal structure of the action that contributes to the trace anomaly, which is entirely reproduced, within the local description, by two auxiliary scalar fields, seems to indicate that the effective dynamics of the coupling between gravity and matter might be controlled, at least in part, by these degrees of freedom. As we have just mentioned, however, this point requires a dedicated study and for this specific reason our conclusions remain open ended.

Our computation, however, being general, allows also the identification of other massless contributions to the effective action which are surely bound to play a role in the physical S-matrix. They appear in form factors such as $\Phi_2$ (Eq. \ref{Phi2}) 
and $\Phi_3$ (Eq. \ref{Phi3}) which do not contribute to the trace, but are nevertheless part of the 1-loop effective action mediated by the triangle graph.  

There are also some other comments, at this point, which are in order. Notice that while the isolation of the pole in the fermion sector indeed requires a massless fermion limit, as obvious from the structure of $\Gamma_q$, the other gauge invariant sector, described by $\Gamma_g$, is obviously not affected by this limit, being the corresponding form factors mass independent. This obviously does not imply necessarily that the gluon pole, which survives the extrapolation to the massless limit, is coupled in the physical S-matrix.  

Building on considerations of this nature, in particular on the possible significance of massless effective degrees of freedom,
the role of the trace anomaly in establishing the effective interaction of gravity with matter has been reconsidered \cite{Urban:2009vy, Urban:2009yg}. The explicit goal of this approach has been to trying to bypass the existing hierarchy problem between the value of the expected vacuum energy density ($\rho \sim (10^{-3} \textrm{eV})^4$), well-described by a cosmological constant, and the Planck mass $(\rho\sim M_P^4)$, which is a fundamental issue in contemporary cosmology that has not found yet a convincing explanation. In fact, it has been known for a long time that free massless particles contribute to the anomaly by an insignificant amount ($T^\mu_\mu\sim H_0^4$), proportional to the fourth power of the current Hubble rate, which is far too small as a value to solve the dark energy problem, due to the fact that we are living in a flat universe. However, it has been suggested that this small value for the vacuum energy density, originally attributed to the anomaly, could be raised to the expected one if the gravitational effective action is characterized by some effective nonlocality. In this case the contribution due to the trace anomaly could be modified as \cite{Klinkhamer:2009nn}

\beq
T^\mu_\mu\sim H_0\, \Lambda_{QCD} ^3 \sim (10^{-3} \textrm{eV})^4,
\eeq
where $\Lambda_{QCD}$ is the QCD scale, which is tantalizingly close to the estimated value. While this proposal and similar others are clearly not the only possible solutions of the dark energy problem (similar values of the vacuum energy can be obtained, for instance, using axions misaligned at the electroweak scale \cite{Nomura:2000yk} and in several other ways) they share the positive feature of being characterized by few minimal assumptions. If so, one could envision a solution of the problem of the origin of dark energy
without the need to enlarge the Standard Model spectrum with yet unknown particles and symmetries.  Crucial, in these types of approaches, appears to be the role played by the effective scalar fields in the anomalous effective action, which are present in the local formulation of Riegert's action, together with their possible boundary conditions.

\section{Conclusions}
One of the standing issues of the anomalous effective action describing the interaction of a non-abelian theory to gravity is a test of its consistency with the standard perturbative approach. Thus, variational solutions of the effective action controlled by the trace anomaly should be reproduced by the perturbative expansion. Building on previous analysis in QED, here we have shown that also in the non-abelian case there is a perfect match between the two approaches. This implies that the interaction of gravity with a non-abelian gauge theory, mediated by the trace anomaly, indeed can be reformulated in terms of auxiliary scalar degrees of freedom, in analogy to the abelian case. We have proven this result by an explicit computation. Our findings indicate that
this feature is typical of each gauge invariant subsector of the non-abelian $TJJ$ amplitude, a result which is likely to hold also for singularities of higher order. These are expected to be present in correlators with a larger number of energy momentum insertions.  We hope to return with a more detailed discussion of the role of the massless singularities 
- which have been found both in the trace and in the traceless part - of the $TJJ$ vertex in the near future.

 \centerline{\bf Acknowledgements}
We are grateful to Emil Mottola for discussions and to the Cern Theory Group for hospitality. This work is supported in part  by the European Union through the Marie Curie Research and Training Network ``Universenet'' (MRTN-CT-2006-035863).

\begin{appendix}
\section{Appendix. Feynman rules}
\label{rules}
The Feynman rules used throughout the paper are collected here
\begin{itemize}
\item {Graviton - fermion - fermion vertex}
\\
\\
\bmi{95pt}
\includegraphics[scale=1.0]{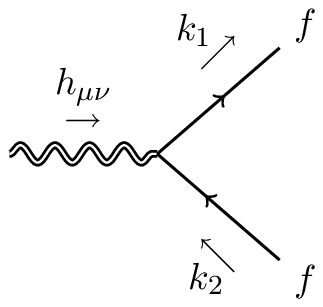}
\emi
\bmi{70pt}
\bann
&=& - i \, \frac{\kappa}{2} \, V^{\prime}_{\mu\nu}(k_1,k_2) \nn \\
&=& - i \, \frac{\kappa}{2} \, \left\{ \frac{1}{4} \left[\gamma_\mu (k_1 + k_2)_\nu
+\gamma_\nu (k_1 + k_2)_\mu \right] - \frac{1}{2} g_{\mu \nu}
[\gamma^{\lambda}(k_1 + k_2)_{\lambda} - 2 m]  \right\} \nn \\
\eann
\emi
\bea
\label{VGff}
\eea
\item{Graviton - gluon - gluon vertex}
\\ \\
\bmi{110pt}
\includegraphics[scale=1.0]{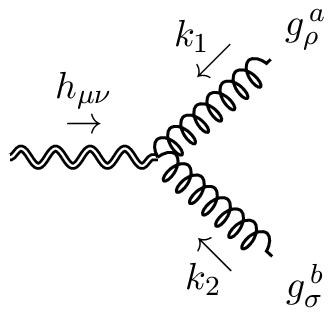}
\emi
\bmi{100pt}
\bann
&=& - i \, \frac{\kappa}{2} \, \delta_{a b} \, V^{Ggg}_{\mu\nu\rho\sigma}(k_1,k_2) \nn \\
 &= &- i \, \frac{\kappa}{2} \, \delta_{a b} \left\{ k_1\cdot k_2 \, C_{\mu\nu\rho\si} + D_{\mu\nu\rho\si}(k_1,k_2) + \frac{1}{\xi} \, E_{\mu\nu\rho\si}(k_1,k_2)  \right\}
\eann
\emi
\bea
\eea
\item{Graviton - ghost - ghost vertex}
\\ \\
\bmi{110pt}
\includegraphics[scale=1.0]{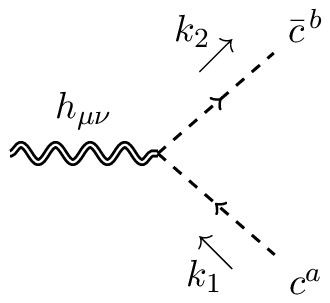}
\emi
\bmi{70pt}
\bann
& = & - i \,  \frac{\kappa}{2}\,  \delta^{a b} \, C_{\mu\nu\rho\sigma} \, k_{1\,\rho} \, k_{2\,\sigma}
\eann
\emi
\bea
\eea
%
%
\item{Graviton - fermion - fermion - gauge boson vertex}
\\ \\
\bmi{110pt}
\includegraphics[scale=1.0]{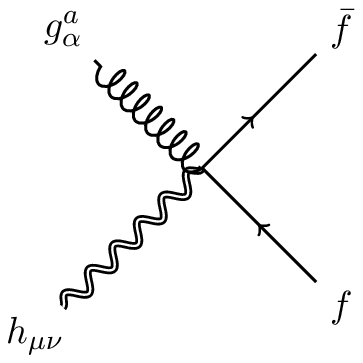}
\emi
\bmi{70pt}
\bann
&=& i g \, \frac{\kappa}{2} \,T^a\, W^{\prime}_{\mu\nu\alpha}
= i g \, \frac{\kappa}{2} \, T^a \left\{ -\frac{1}{2} (\gamma_\mu \, g_{\nu\alpha}
+\gamma_\nu \, g_{\mu\alpha}) +  g_{\mu \nu} \, \gamma_{\alpha} \right \}
\eann
\emi
\bea
\label{WGffg}
\eea
%
%
%
\item{Graviton - gluon - gluon - gluon  vertex}
\\ \\
\bmi{110pt}
\includegraphics[scale=1.0]{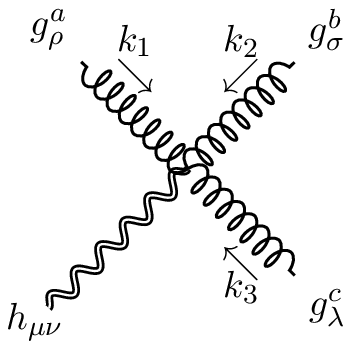}
\emi
\bmi{110pt}
\begin{eqnarray*}
&=&- g \frac{\kappa}{2} f^{a b c} V^{Gggg}_{\mu\nu\rho\sigma\lambda}(k_1,k_2,k_3) \nn \\
&=&  - g \frac{\kappa}{2} f^{a b c} \left\{ C_{\mu\nu\rho\sigma}(k_1-k_2)_{\lambda} + C_{\mu\nu\rho\lambda}(k_3-k_1)_{\sigma}   \right. \nn \\
&& \hspace{2.5cm}  + \left.  C_{\mu\nu\sigma\lambda}(k_2-k_3)_{\rho} + F_{\mu\nu\rho\sigma\lambda}(k_1,k_2,k_3)  \right\}
\hspace{1.7cm}
\end{eqnarray*}
\emi
\bea
\eea
%
%
\item{Graviton - ghost - ghost - gauge boson vertex}
\\ \\
\bmi{110pt}
\includegraphics[scale=1.0]{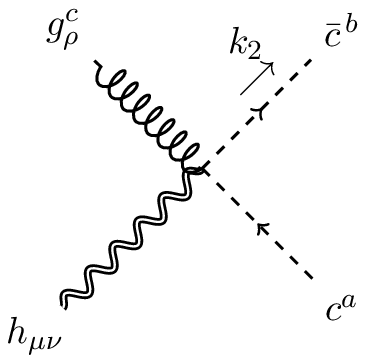}
\emi
\bmi{70pt}
\bann
&=& \frac{\kappa}{2} \, g \, f^{a b c} \, C_{\mu\nu\rho\sigma} \, k_{2}^{\sigma}
\eann
\emi
\bea
\eea
%
%
\bea
&& C_{\mu\nu\rho\sigma} = g_{\mu\rho}\, g_{\nu\sigma}
+g_{\mu\sigma} \, g_{\nu\rho}
-g_{\mu\nu} \, g_{\rho\sigma}\,
\\
&& D_{\mu\nu\rho\sigma} (k_1, k_2) =
g_{\mu\nu} \, k_{1 \, \sigma}\, k_{2 \, \rho}
- \biggl[g^{\mu\sigma} k_1^{\nu} k_2^{\rho}
  + g_{\mu\rho} \, k_{1 \, \sigma} \, k_{2 \, \nu}
  - g_{\rho\sigma} \, k_{1 \, \mu} \, k_{2 \, \nu}
  + (\mu\leftrightarrow\nu)\biggr]\, \\
&& E_{\mu\nu\rho\sigma} (k_1, k_2) = g_{\mu\nu} \, (k_{1 \, \rho} \, k_{1 \, \sigma}
+k_{2 \, \rho} \, k_{2 \, \sigma} +k_{1 \, \rho} \, k_{2 \, \sigma})
-\biggl[g_{\nu\sigma} \, k_{1 \, \mu} \, k_{1 \, \rho}
+g_{\nu\rho} \, k_{2 \, \mu} \, k_{2 \, \sigma}
+(\mu\leftrightarrow\nu)\biggr]\ , \nn  \\ \\
&& F_{\mu\nu\rho\sigma\lambda} (k_1,k_2,k_3) =
g_{\mu\rho} \,  g_{\sigma\lambda} \, (k_2-k_3)_{\nu}
+g_{\mu\sigma} \, g_{\rho\lambda} \, (k_3-k_1)_{\nu}
+g_{\mu\lambda} \, g_{\rho\sigma}(k_1-k_2)_{\nu}
+ (\mu\leftrightarrow\nu) \nn \\
\eea
\end{itemize}
\section{Appendix. Scalar integrals}
\label{scalars}
We collect in this appendix all the scalar integrals involved in this computation. To set all our conventions, we start with the definition of the one-point function, or massive tadpole $\mathcal  A_0 (m^2)$, the massive bubble $\mathcal B_0 (s, m^2) $  and  the massive three-point function $\mathcal C_0 (s, s_1, s_2, m^2)$
\bea
\mathcal A_0 (m^2) &=& \frac{1}{i \pi^2}\int d^n l \, \frac{1}{l^2 - m^2}
= m^2 \left [ \frac{1}{\bar \eps} + 1 - \log \left( \frac{m^2}{\mu^2} \right )\right],\\
 \mathcal B_0 (k^2, m^2) &=&  \frac{1}{i \pi^2} \int d^n l \, \frac{1}{(l^2 - m^2) \, ((l - k )^2 - m^2 )} \nn \\
 &=& \frac{1}{\bar \eps} + 2 - \log \left( \frac{m^2}{\mu^2} \right ) - a_3 \log \left( \frac{a_3+1}{a_3-1}\right), \\
\mathcal C_0 (s, s_1, s_2, m^2) &=&
 \frac{1}{i \pi^2} \int d^n l \, \frac{1}{(l^2 - m^2) \, ((l -q )^2 - m^2 ) \, ((l + p )^2 - m^2 )} \nn \\
&=&- \frac{1}{ \sqrt \sigma} \sum_{i=1}^3 \left[Li_2 \frac{b_i -1}{a_i + b_i}   - Li_2 \frac{- b_i -1}{a_i - b_i} + Li_2 \frac{-b_i +1}{a_i - b_i}  - Li_2 \frac{b_i +1}{a_i + b_i}
   \right],
\label{C0polylog}
\eea
with
\bea
a_i = \sqrt {1- \frac{4 m^2}{s_i }} \qquad \qquad
b_i = \frac{- s_i + s_j + s_k }{\sqrt{ \sigma}},
\eea
where $s_3=s$ and in the last equation $i=1,2,3$ and $j, k\neq i$. \\
The one-point and two-point functions  written before in  $n=4 - 2 \, \eps$ are divergent in dimensional regularization with the singular parts given by
\bea
\mathcal A_0 (m^2) ^{sing.}  \rightarrow  \frac{1}{\bar \eps} \, m^2,  \qquad \qquad
\mathcal B_0 (s, m^2) ^{sing.}  \rightarrow  \frac{1}{\bar \eps} ,
\eea
with
\bea
\frac{1}{\bar \eps} = \frac{1}{\eps} - \g - \ln \pi
\label{bareps}
\eea
We use two finite combinations of scalar functions given by
\bea
&&  \mathcal B_0 (s, m^2) \, m^2 - \mathcal A_0 (m^2) =  m^2 \left[ 1 - a_3 \log \frac{a_3 +1}{a_3 - 1}  \right] , \\
&& \mathcal D_i \equiv \mathcal D_i (s, s_i,  m^2) =
\mathcal B_0 (s, m^2) - \mathcal B_0 (s_i, m^2) =  \left[ a_i \log\frac{a_i +1}{a_i - 1}
- a_3 \log \frac{a_3 +1}{a_3 - 1}  \right] \qquad i=1,2.
\label{D_i}
\nn \\
\eea
The scalar integrals $ \mathcal C_0 (s, 0,0,m^2) $ and $\mathcal D (s, 0, 0, m^2)$ are the $ \{ s_1 \rightarrow 0$, $s_2 \rightarrow 0 \} $ limits of the generic functions $\mathcal C_0(s,s_1,s_2,m^2)$ and $\mathcal D_1(s,s_1,m^2)$
\bea
\mathcal C_0 (s, 0,0,m^2) &=& \frac{1}{2 s} \log^2 \frac{a_3+1}{a_3-1}, \\
\mathcal D (s, 0, 0, m^2) &=& \mathcal D_1 (s,0,m^2)= \mathcal D_2(s,0,m^2) =
  \left[ 2 - a_3 \log \frac{a_3+1}{a_3-1}\right].
\eea
The singularities in $1/\bar \eps$ and the dependence on the renormalization scale $\mu$ cancel out when taking into account the difference of two functions $\mathcal B_0$, so that the $\mathcal D_i$'s  are well-defined; the three-point master integral is convergent.

The renormalized scalar integrals  in the modified minimal subtraction scheme named $\overline{MS}$ are defined as
\bea
\mathcal B_0^{\overline{MS}}(s,0) &=&  2 - L_s,
\label{B0masslessOS}\\
\mathcal B_0^{\overline{MS}}(0,0) &=&  \frac{1}{\omega},\\
\mathcal C_0(s,0,0,0) &=& \frac{1}{s}\left[ \frac{1}{\omega^2} + \frac{1}{\omega} L_s + \frac{1}{2} L_s^2 - \frac{\pi^2}{12}  \right],
\label{C0masslessOS}
\eea
where
\bea
L_s \equiv \log \left( - \frac{s}{\mu^2} \right ) \qquad \qquad s<0.
\eea
We have set  the space-time dimensions to $n = 4 + 2 \omega$ with $\omega > 0$. The $1/\omega$ and $1/\omega^2$ singularities in Eqs.~(\ref{B0masslessOS}) and (\ref{C0masslessOS}) are infrared divergencies due to the zero mass of the gluons.

\end{appendix}

\end{document}